\documentclass[aps,prx,twocolumn,superscriptaddress,nofootinbib,showkeys]{revtex4-1}

\usepackage{graphicx}
\usepackage{dcolumn}
\usepackage{bm}
\usepackage[utf8]{inputenc}
\usepackage{textcomp}
\usepackage{amsmath}
\usepackage[mathlines]{lineno}
\usepackage{siunitx}
\usepackage{hyperref}
\usepackage[dvipsnames]{xcolor}
\usepackage[final]{pdfpages}
\usepackage{pgffor}

\makeatletter
\AtBeginDocument{\let\LS@rot\@undefined}					
\makeatother

\hypersetup{
    colorlinks=true,
    citecolor=blue,
    urlcolor=blue,
    linkcolor=black
}

\begin{document}

\title{TMDC Resonators for Second Harmonic Signal Enhancement}

\author{Sebastian Busschaert}
\author{Ren{\'e} Reimann}
\author{Moritz Cavigelli}
\author{Ronja Khelifa}
\author{Achint Jain}
\author{Lukas Novotny}

\affiliation{Photonics Laboratory, ETH Zürich, CH-8093 Zürich, Switzerland 
}

\date{\today}

\begin{abstract}
In addition to their strong nonlinear optical response, transition metal dichalcogenides (TMDCs) possess a high refractive index in the visible and infrared regime. Therefore, by patterning those TMDCs into dielectric nanoresonators, one can generate highly confined electromagnetic modes. Controlled fabrication of TMDC nanoresonators does not only enhance the material's intrinsic nonlinear response, but also allows for spatially shaping the emission via nanoresonator arrays. Here we fabricate patterned WS\textsubscript{2} disks that support a high internal resonant electric field and show strong enhancement of second harmonic (SH) generation in the visible regime. In addition, we assemble the WS\textsubscript{2} disks in arrays to spatially direct the coherent SH  emission, in analogy to phased array antennas. Finally, we investigate and discuss drastic differences in the areal emission origin and intensity of the measured SH signals, which we find to depend on material variations of the used bulk WS\textsubscript{2}.\end{abstract}

\maketitle

In their few-layer form, transition metal dichalcogenides (TMDCs) exhibit high nonlinear optical responses~\cite{Kumar13,Malard13,Woodward16}. 
However, the small thickness of the few-layer form results in a low interaction volume. 
Attempts to increase the interaction volume have led to several studies in which the intrinsic TMDC nonlinearity has been enhanced via light-confining structures. 
Until now, these studies have been focused exclusively on coupling few-layer TMDCs to various external systems, such as plasmonic structures~\cite{Wang18,Chen18,Hu19}, microcavities~\cite{Day16,Yi16}, silicon photonic devices~\cite{Fryett16,Chen17} or dielectric nanowires~\cite{Li19}. 
Here, we utilize a TMDC material itself as optical resonators to enhance their inherent nonlinearity. 
Our approach is highly promising, as TMDCs exhibit a high refractive index~\cite{Yim14,Li14}, making them ideal candidates to be patterned into high-index dielectric resonators. 
The utilization of a stand-alone TMDC resonator does not only result in a higher damage threshold compared to plasmonic systems, but also reduces the complexity in fabrication compared to multi-material photonic systems.
High-index dielectric resonators can support a variety of electromagnetic modes~\cite{Krasnok12,Kuznetsov16}. 
In order to boost the intrinsic nonlinear response of the patterned material, the anapole mode, which is neither a pure electric or magnetic dipole mode, has attracted the focus of the scientific community~\cite{Yang19,Savinov19,Tian19}. 
This mode exhibits low energy losses compared to other modes (e.g. dipolar mode) and the high field concentration inside the resonator presents an ideal playground for enhancing material nonlinearities. Most recent efforts to fabricate anapole resonators have focused on the use of III-V materials~\cite{Timofeeva18}, Silicon~\cite{Mirosh15,Baranov18} and Germanium~\cite{Grinblat16,Grinblat17}.
Nevertheless, the same design criteria can be used for TMDC resonators~\cite{Verre19}. 
In this work we present patterned WS\textsubscript{2} disks that exhibit anapole-like resonance behavior which resonantly enhances the intrinsic 
second harmonic (SH) generation. 
By exciting several disks in an array we are also able to spatially direct the nonlinear emission and to infer the areal origin of the SH signal.\\
\begin{figure}[!ht] 
\centering
\includegraphics[width=\columnwidth]{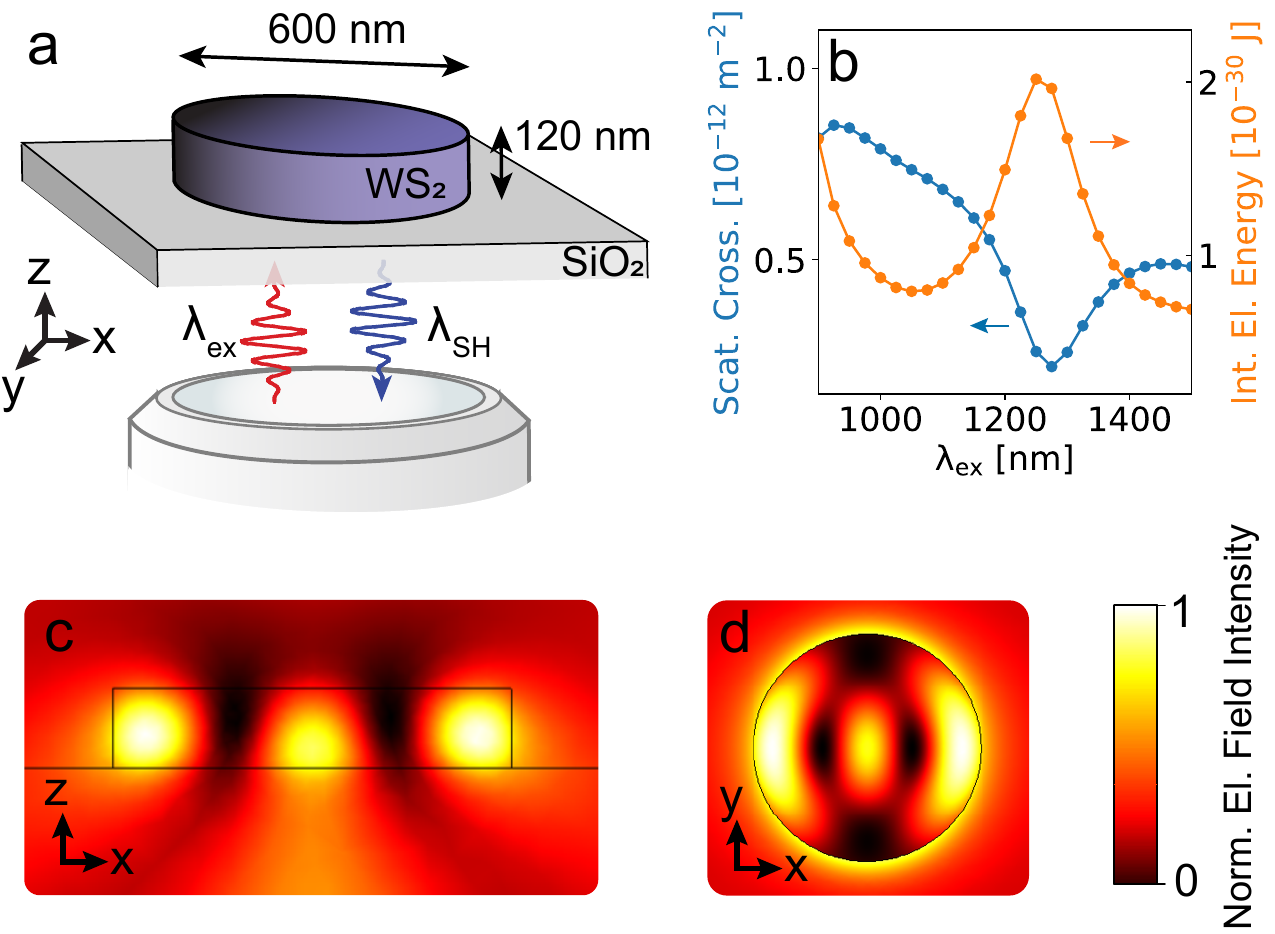}
\caption{Electromagnetic resonance of a WS\textsubscript{2} disk on a glass substrate. (a) Geometry and dimensions of the nanoresonator. (b) Electrodynamic simulation via finite element method. The blue curve shows the scattering cross-section and the orange curve the internal electric energy as a function of the excitation wavelength. 
(c,d) Normalized intensity cross-sections of the disk for excitation at $1250\,{\rm nm}$. The incoming linear polarization is chosen along the $y$ axis. The exciting plane wave electric field has a value of $1\,{\rm V/m}$.}
\label{fig:anap}
\end{figure}
First, we discuss the design and the resonance conditions of our resonators. 
The resonance wavelength, as for other antenna structures, is determined by the resonator geometry~\cite{Novotny12}. 
In Fig.~\ref{fig:anap} we demonstrate a finite-element simulation of a WS\textsubscript{2} disk, that shows anapole-like behavior at an excitation wavelength of $1250\,{\rm nm}$. 
The hallmarks of the mode's resonance, apart from its distinct near field pattern, are a dip in the far field scattering cross-section and an increase in the internal electric energy. 
The latter is of particular interest for enhancing nonlinear effects. 
If we choose the excitation wavelength to be at the anapole resonance, the resulting enhancement should be directly observable in the third order nonlinearity, as third harmonic (TH) generation has been shown to be more efficient in bulk than in few-layer TMDCs~\cite{Wang14,Khorasani18,Balla18}. 
Until now, for second order effects the conversion efficiency is reported to be significantly weaker for bulk than for a few material layers~\cite{Li13,Zhao16}. \\
\begin{figure}[!ht] 
\centering
\includegraphics[width=\columnwidth]{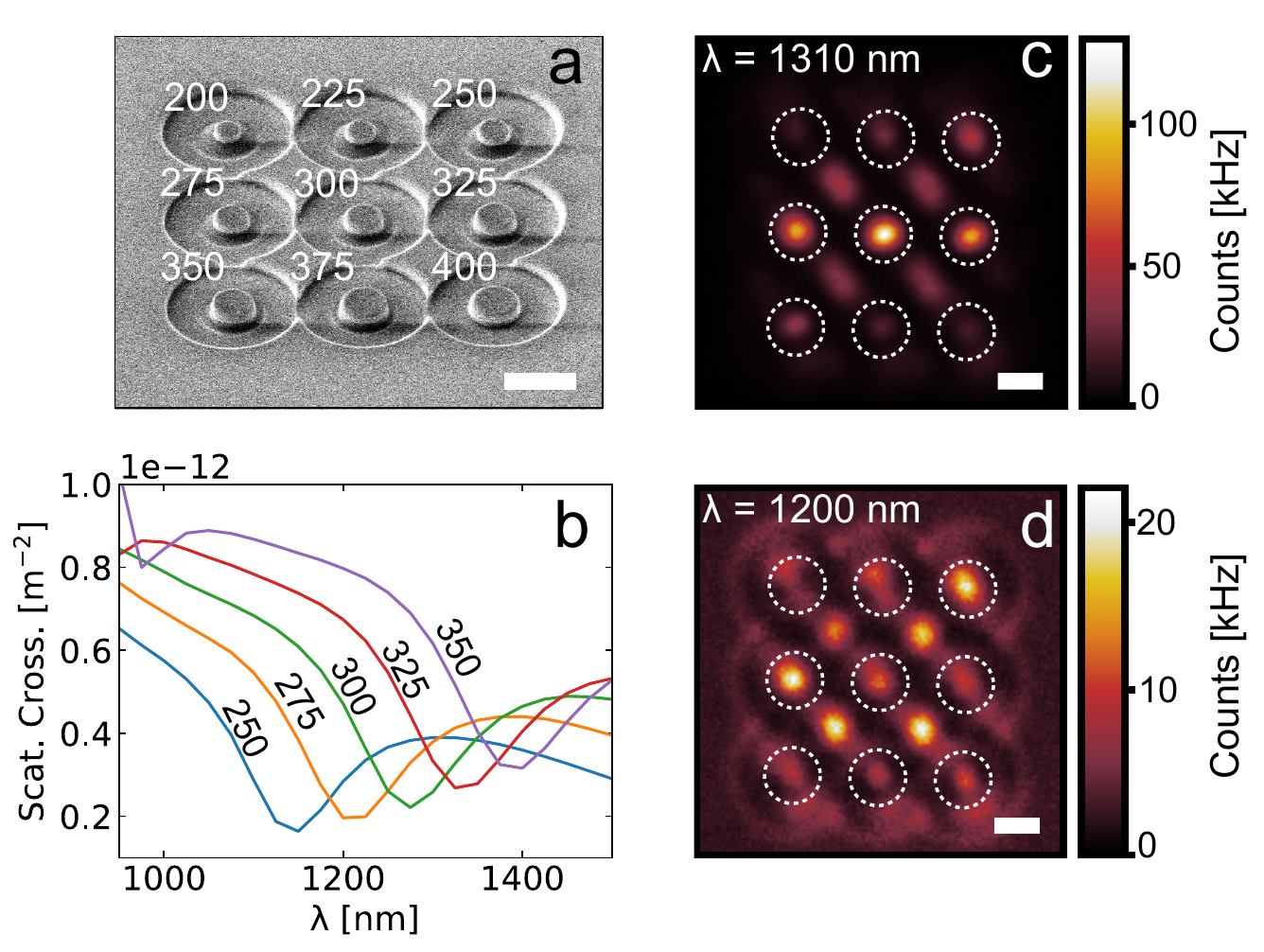}
\caption{Third-harmonic (TH) generation by single WS\textsubscript{2} disks. (a) SEM image of fabricated resonators. The values are the design radii $R$ in \SI{}{\nano\meter}. The scale bar is \SI{1}{\micro\meter}. (b) Simulated scattering cross-sections for disks with varying radii. (c,d) TH intensity maps of disks according to (a) for excitation wavelengths of (c) \SI{1310}{\nano\meter} and (d) \SI{1200}{\nano\meter}. The scale bars are \SI{1}{\micro\meter}, the disk positions are marked with dotted circles.}
\label{fig:TH}
\end{figure}
We use the results of our simulations to obtain a target parameter space and fabricate WS\textsubscript{2} disks accordingly. 
For that, we exfoliate bulk WS\textsubscript{2} on top of a glass substrate and pattern disks via focused ion beam (FIB) milling (for details see Supporting Information). 
In Fig.~\ref{fig:TH}a we display a scanning electron microscope  (SEM) image of individual disks, with radii spanning from \SI{200}{} to \SI{400}{\nano\meter} with increments of \SI{25}{\nano\meter}. 
We use TH generation as a characterization tool to test resonant field enhancement inside the disks.
To this end, we scan over the resonators with \SI{1310}{nm} excitation light and spectrally filter the emitted light via a dichroic beamsplitter and optical filters.
The TH emission at \SI{437}{nm} is collected in the resulting transmission window at wavelengths below \SI{561}{nm}. 
According to simulated scattering curves (see Fig.~\ref{fig:TH}b), a disk with a radius of around \SI{300}{\nano\meter} should exhibit the highest intensity enhancement by the anapole mode, as the resonance of this disk is closest to the chosen laser excitation wavelength. 
The experimental TH intensity map is displayed in Fig.~\ref{fig:TH}c. 
Indeed the disk with a radius of \SI{300}{\nano\meter} exhibits the highest TH signal, whereas disks with smaller and larger radii show decreased TH emission (for spectra see Supporting Information). 
We thus indirectly map out the resonance profile of the enhancing mode. 
To confirm the wavelength dependent resonance behavior of our disks, we tune the excitation wavelength to \SI{1200}{\nano\meter}. 
In the resulting TH intensity map (Fig.~\ref{fig:TH}d) we now observe that disks with smaller radii are scattering stronger. 
In particular the disk with a \SI{275}{\nano\meter} radius shows the highest signal, again agreeing with the simulated scattering curves.
The decrease in TH intensity for the lower excitation wavelength could be attributed to the higher absorption of WS\textsubscript{2} at the corresponding TH wavelength of \SI{400}{\nano\meter} and the lower detection efficiency of the used detector. 
The observed TH scattering intensity dependence on the disk radius and on the excitation wavelength, combined with the discussed simulations, indicates that we have successfully built resonators that exhibit anapole-like behavior. 


\begin{figure*}[!ht] 
\centering
\includegraphics[width=\textwidth]{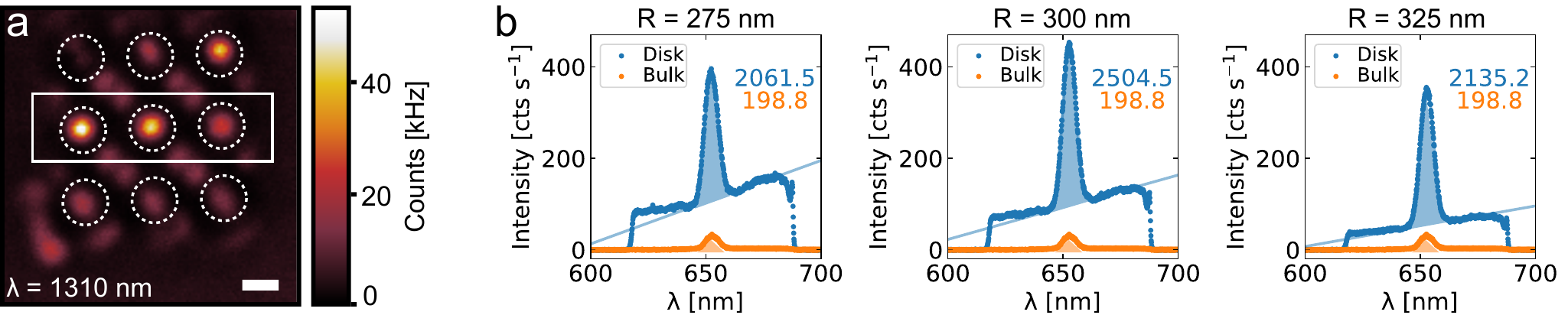}
\caption{SH intensity measurement of single WS\textsubscript{2} disks and of unstructured WS\textsubscript{2} bulk material next to disks. (a) Measured SH intensity map. The disk layout is identical to Fig.~\ref{fig:TH}a but it's a different device. The scale bar is \SI{1}{\micro\meter}. (b) Spectra of the disks (blue curves) within the white rectangle in (a), from left to right. The width of the spectral peaks is defined by the excitation laser pulse duration, see Methods. The orange curve shows a spectrum taken from the bulk WS\textsubscript{2} next to the disks. The numbers next to the SH peaks indicate the respective SH signal strengths $S$ (light blue and light orange shaded areas).}
\label{fig:SH}
\end{figure*}

Having established the functionality of our nanoresonators, we now turn our attention to the main focus of this work, i.e. enhanced SH generation in resonant WS\textsubscript{2} structures. First, we aim to develop a basic understanding of the physical processes at play.
To theoretically describe the resonant SH emission we have to account for the overlap of the relevant electric fields inside the resonator. Those relevant fields are $\bm{E}_{\rm{SH}}(\bm{r})$ at the SH and $\bm{E}_{\rm{ex}}(\bm{r})$ at the fundamental wavelength. 
Following references~\cite{Buckley14,Lin16} we express the emitted power at the SH wavelength as
\begin{equation}\label{eq:overlap}
    \begin{split}
    P_{\rm{SH}} &= \frac{64\lambda_{\rm{ex}}}{\pi c } (\eta_{\rm{inc}}P_{\rm{inc}})^2 \\
    &\times\left|\frac{\int \epsilon_0\chi^{(2)}(\bm{r})\bm{E}_{\rm{SH}}(\bm{r})\bm{E}_{\rm{ex}}(\bm{r})^2 {\rm d}V}{\int\epsilon_{\rm{ex}}|\bm{E}_{\rm{ex}}(\bm{r})|^2{\rm d}V \sqrt{\int\epsilon_{\rm{SH}}|\bm{E}_{\rm{SH}}(\bm{r})|^2{\rm d}V}}\right|^2 ,
    \end{split}
\end{equation}
where $V$ is the volume, $\epsilon_0$ is the vacuum permittivity, $c$ is the vacuum speed of light, $\lambda_{\rm{ex}}$ is the fundamental wavelength in bulk, $\epsilon_{\rm{ex}}$, $\epsilon_{\rm{SH}}$ are the absolute permittivities of WS\textsubscript{2} at the respective wavelengths, $P_{\rm inc}$ is the incoming power of the excitation laser, $\eta_{\rm inc}$ is the coupling efficiency of $P_{\rm inc}$ to the fundamental resonator field $\bm{E}_{\rm{ex}}(\bm{r})$, and $\chi^{(2)}(\bm{r})$ is the second order susceptibility of WS\textsubscript{2}.
According to~\cite{Yan18,Gigli19} we can expand both the fundamental and SH fields in the WS\textsubscript{2} disks in terms of quasinormal modes as
\begin{gather}
  \bm{E}_{\rm{ex}}(\bm{r}) = \sum_{i=1}^{N} \alpha_i(\lambda_{\rm{ex}})\widetilde{\bm{E}}_i^{(\lambda_{\rm ex})}(\bm{r})
  , \\
  \bm{E}_{\rm{SH}}(\bm{r}) = \sum_{i=1}^{M} \alpha'_i(\lambda_{\rm{SH}})\widetilde{\bm{E}}_i^{(\lambda_{\rm SH})}(\bm{r}),
\end{gather}{}
where $\lambda_{\rm SH}$ is the SH wavelength in bulk, $\alpha_i$ and $\alpha'_i$ are the modal excitation coefficients, which include the respective quality factors and coupling efficiencies, and $\widetilde{\bm{E}}_i^{(\lambda)}(\bm{r})$ are the quasinormal modes of the resonator at wavelength $\lambda$.
For instance, the resulting field intensity at the fundamental wavelength $\bm{E}_{\rm{ex}}(\bm{r})$, displayed in Fig.~\ref{fig:anap}, can be written as the sum of an electric dipole mode and a toroidal dipole mode \cite{Yang19,Savinov19,Tian19}.
Please note, that $\alpha_i$ reflects the modal overlap between the incoming field of the excitation laser and $\widetilde{\bm{E}}_i^{(\lambda_{\rm SH})}(\bm{r})$, whereas $\alpha'_i$ reflects the modal overlap between the excitation field $\bm{E}_{\rm{ex}}(\bm{r})$ and $\widetilde{\bm{E}}_i^{(\lambda_{\rm ex})}(\bm{r})$~\cite{Yan18,Gigli19}.\\ 
Having discussed the theoretical basis of SH generation in nanoresonators, we now study the measured SH response of our WS\textsubscript{2} nanoresonators.
Figure~\ref{fig:SH}a shows the SH intensity map of the disk array. 
Here we use a new device, but the disk layout is identical to Fig.~\ref{fig:TH}a, and the device shows the same TH behaviour as displayed in Fig.~\ref{fig:TH}c.
At first sight, it appears that the disk with a radius of \SI{275}{\nano\meter} exhibits the highest SH signal.
However, when studying the spectra of the respective disks within the white box (see Fig.~\ref{fig:SH}b) one finds that the radius of the disk with resonant SH generation is equivalent to the radius of the disk with resonant TH generation in Fig.~\ref{fig:TH}c. 
This is clear proof that the observed resonance behavior is indeed given by the resonance at the fundamental wavelength. 
The discrepancy between the SH intensity map and the spectra results from the fact, that the map is recorded with an avalanche photo diode (APD) that integrates all the light that passes the optical filters. 
Therefore unwanted incoherent background is recorded in addition to the coherent SH signal. The smaller background level for larger disks in Fig.~\ref{fig:SH}b could stem from higher absorption due to the larger disk volume of large disks. 
We next compare the SH signal from the \SI{300}{nm} radius disk $S_{\text{disk}}$ with the SH signal originating from the unstructured bulk next to the disks $S_{\text{bulk}}$.
We find a SH enhancement factor
\begin{equation}
\label{eq:enhancement}
    f_{\rm SH} = \frac{S_{\text{disk}}}{S_{\text{bulk}}}\frac{A_{\text{bulk}}}{A_{\text{disk}}} = \frac{S_{\text{disk}}}{S_{\text{bulk}}}\frac{r_{\text{laser}}^2}{r_{\text{disk}}^2} = 50 \pm 5    ,
\end{equation}
where we account for the differences in excited area $A$ of the bulk measurement [given by the excitation beam's $1/e^2$ intensity radius $r_{\text{laser}}=(\SI{600}{} \pm \SI{30}{})\, {\rm nm}$] and the disk measurement (given by the disk radius $r_{\text{disk}}=\SI{300}{nm}$). 
Thus, by exploiting a resonance at the excitation wavelength, we successfully demonstrate a significant SH signal enhancement.
\begin{figure*}[!ht]
\centering
\includegraphics[width=\textwidth]{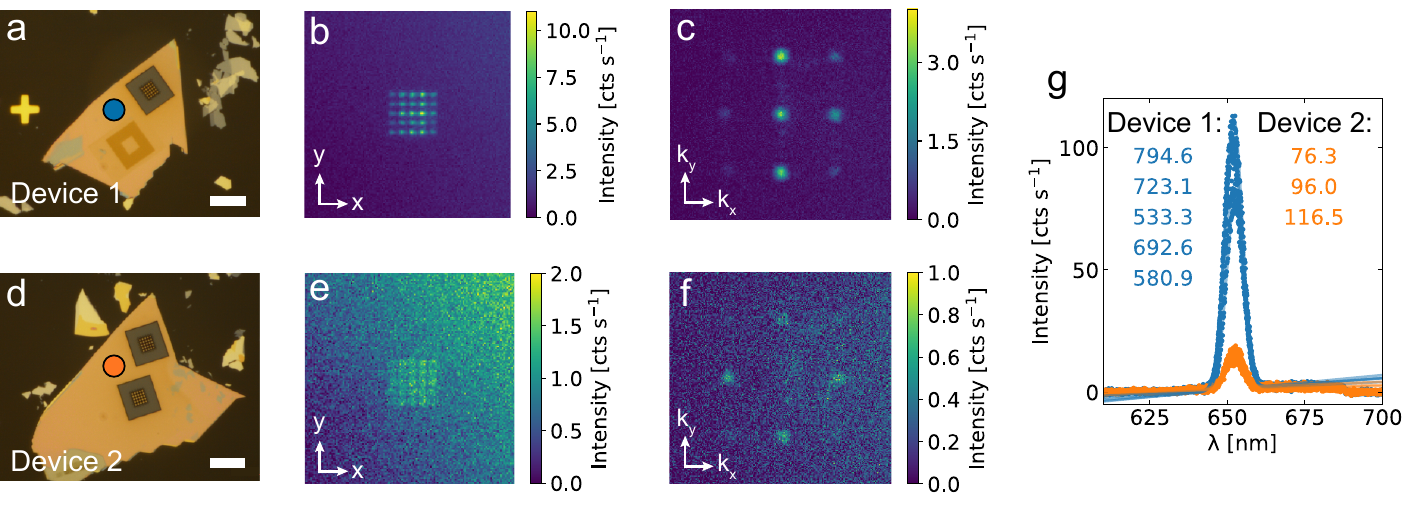}
\caption{SH measurement of WS\textsubscript{2} resonator arrays and bulk material. Both arrays consist of 5x5 identical disks with a WS\textsubscript{2} height of \SI{123}{nm} and a design radius of \SI{300}{nm}. (a) Reflection microscope image of WS\textsubscript{2} resonator array (device~1). The scale bar is \SI{10}{\micro\meter}. (b,c) Real and back focal plane images of the emitted SH radiation of device~1. (d) Reflection microscope image of WS\textsubscript{2} resonator arrays, including array (device~2) with same parameters as in (a). The scale bar is \SI{10}{\micro\meter}. (e,f) Real and back focal plane images of the emitted SH radiation of device~2. The background gradient is due to residual stray light. (g) SH response of the WS\textsubscript{2} device bulk materials. The respective measurement positions are marked with blue and orange circles in (a,d). The SH responses were confirmed to be homogeneous over the whole bulk material. The numbers indicate the areas of Gaussian fits to the SH curves. The differences for one material are due to rotated incoming linear polarizations (between $0$ and $90^\circ$). The excitation wavelength is \SI{1310}{nm} for all displayed measurements.}
\label{fig:SH_array}
\end{figure*}

After having investigated the resonance behavior of single resonators, we now study the collective SH radiation of a closely spaced disk array. 
We use the same WS\textsubscript{2} thickness (\SI{123}{\nano\meter}) as for the single disk measurements (6\% uncertainty). 
We choose the radius to be \SI{300}{\nano\meter}, as this leads to the strongest SH signal for an excitation wavelength of \SI{1310}{\nano\meter}.
The spacing between resonators is chosen to be \SI{780}{\nano\meter}. 
In order to illuminate a large area and thereby excite the whole disk array, we focus the incoming excitation of \SI{1310}{\nano\meter} onto the back focal plane (BFP) of the 1.3 NA oil immersion objective used for imaging and signal collection. 
This results in a beam spot with a $1/e^2$ intensity diameter of $\sim$ \SI{10}{\micro\meter}.
The average power on the sample is \SI{1.1}{\milli\watt}. 
For SH detection we image the spectrally filtered real space of the sample, as well as the BFP. 
Figure \ref{fig:SH_array}a and d show microscope images of two devices (devices 1 and 2) that were identically fabricated. 
The measured intensity maps are also displayed in Fig.~\ref{fig:SH_array} and reveal that the two devices drastically differ in their SH response. 
First, we observe a higher overall signal count in device~1 compared to device~2. 
Second, we find fundamentally different far field radiation patterns for the two devices.
We note that these far field patterns are indicative of the individual antenna radiation patterns ~\cite{Milligan05,Balanis12} and allow us to differentiate between SH emission originating from the bulk or the surface of the disk resonators. 
In Fig.~\ref{fig:SH_array}b (device~1) we observe the inner part of the disks lighting up (bulk emission) whereas in Fig.~\ref{fig:SH_array}e (device~2) we observe the outer part of the disks lighting up (surface emission).  
This difference is even more pronounced in the recorded BFP images. 
Here we record a zeroth order emission lobe (central lobe) in Fig.~\ref{fig:SH_array}c (device~1). 
Since a dipole emission pattern allows for a zeroth order emission lobe, the BFP image of device~1 suggests a dipolar SH emitter that originates from the bulk of the material. 
This emission lobe, however, is absent in Fig.~\ref{fig:SH_array}f (device~2). 
Here, the BFP image reminds of the far field SH pattern of a gold rod array, cf.~\cite{Busschaert19}, where the SH emission originates from the surface of the resonators.
This surface emission suggests opposite dipoles with a phase difference of $\pi$ leading to destructive interference at an emission angle of 0\textdegree. 
We note that the SH emission behavior observed in device~2 is representative of other measured devices, also for devices with slightly varying bulk material thickness (\SI{95}{} to \SI{123}{nm}) or disk radius (\SI{270}{} to \SI{300}{nm}).


We have shown that we have fabricated WS\textsubscript{2} resonator arrays which behave distinctly differently, despite identical fabrication parameters. 
The observed difference in SH emission can be explained by varying second order nonlinear properties ($\chi^{(2)}$) of equally thick (here around \SI{120}{nm}) WS\textsubscript{2} bulk material. 
In order to test this hypothesis, we measure the SH response from the non-structured regions of the WS\textsubscript{2} which forms the material basis of the two presented devices, cf. Fig.~\ref{fig:SH_array}g. 
The two respective regions are marked with circles in Fig.~\ref{fig:SH_array}a,d.
We see a drastic difference in SH signal strength between the two device materials. 
The SH signal of the bulk WS\textsubscript{2} material~1 (material of device~1) is seven times stronger than the signal of the bulk WS\textsubscript{2} material~2 (material of device~2). 
We also tested the polarization dependence by rotating the linear incoming polarization and measuring the respective SH response.
The results are shown in Fig.~\ref{fig:SH_array}g.
The standard deviation over the mean is \SI{14}{\percent} for bulk material 1 and \SI{17}{\percent} for bulk material 2.
We conclude that rotating the angle of the linear polarization leads to a negligible change in signal. 
To exclude the possibility that steps of single-layer WS\textsubscript{2} on the surface of the disks are responsible for this difference in SH yield, we performed SH measurements on other WS\textsubscript{2} bulk materials with measurable single-layer steps.
We found that the signal difference (before and after step) was less than a factor of 2 (see Supporting Information).

\begin{figure*}[!t] 
\centering
\includegraphics[width=\textwidth]{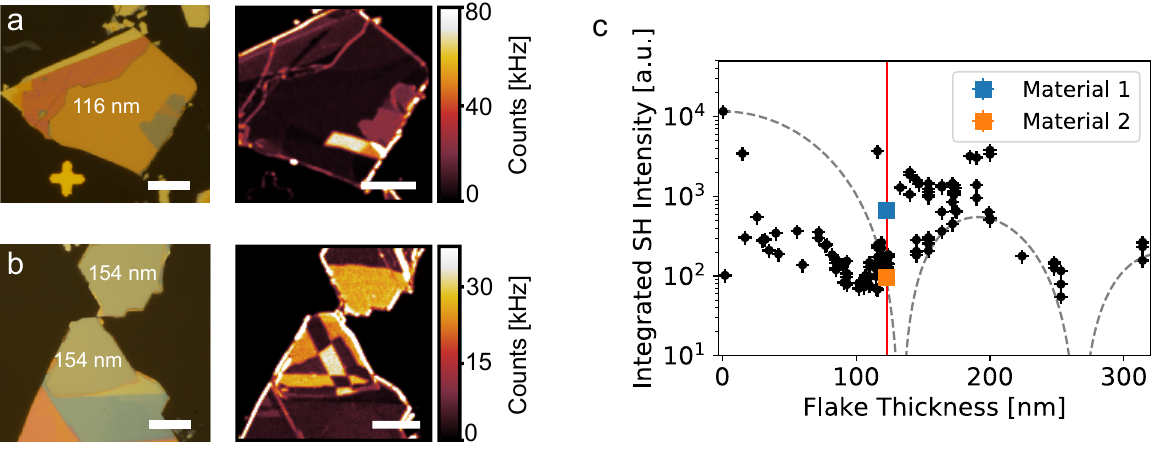}
\caption{SH response study for WS\textsubscript{2} bulk samples of different thicknesses. 
(a,b) Example bulk flakes (left: reflection microscope images) with constant thicknesses (\SI{116}{\nano\meter} and \SI{154}{\nano\meter}, respectively) that reveal bright patches in SH intensity maps (right). 
The scale bars are \SI{10}{\micro\meter}. (c) Measured SH intensities for different bulk thicknesses. The red line marks the thickness of the above discussed bulk materials 1 and 2. The grey dashed line indicates a theoretical curve arising from the phase matching condition for bulk WS\textsubscript{2} crystals. Error bars are due to AFM measurement errors (horizontal) and polarization dependence/power fluctuations (vertical).}
\label{fig:SH_flakes}
\end{figure*}

To get a deeper insight into the bulk material variations of our WS\textsubscript{2} we conduct a statistical study of the material's SH response as function of bulk thickness. 
Our findings are summarized in Fig.~\ref{fig:SH_flakes}. 
In Fig.~\ref{fig:SH_flakes}a and b we demonstrate two samples of WS\textsubscript{2} bulk materials with spatially varying SH responses (see bright patches). 
The samples are left pristine after exfoliation (no further processing). 
We recorded high resolution atomic force microscope (AFM) scans to confirm that these samples are constant in thickness. 
Comparing bright and dark patches within one bulk material sample of constant thickness, leads to a difference in SH intensity of up to one order of magnitude. 
However, we also find that the SH yield depends on the bulk material thickness.
In Fig.~\ref{fig:SH_flakes}c we show the measured SH responses of bulk flakes with different thicknesses. 
We mark the average responses of the materials used for devices 1 and 2 discussed previously in Figs.~\ref{fig:SH_array} (see red line).
Many samples with a thickness of $\sim$ \SI{123}{\nano\meter} have been measured most of which show a SH yield that agrees with the SH yield of device~2.
Based on a simple phase matching model~\cite{Boyd08}, one would expect the integrated SH intensity of bulk WS\textsubscript{2} to follow a sinc-function behavior 
\begin{equation}
    I_{\text{SH}} = I_{\text{ML}} \left | \frac{\sin(\Delta k \times t/2)}{\Delta k \times t/2} \right |^2,
\end{equation}{}
where $I_{\text{ML}}$ corresponds to the integrated SH intensity of a monolayer, $\Delta k = 2k_{\rm ex} - k_{\rm SH}$ is the wave vector mismatch between the bulk SH wave vector $k_{\rm SH} = 2\pi /\lambda_{\rm SH}$ and the bulk excitation wave vector $k_{\rm ex}= 2\pi/\lambda_{\rm ex}$.
The thickness $t$ is weighted with a factor of 1.25 accounting for the high NA objective which increases the effective thickness seen by the light. 
The data shown in Fig.~\ref{fig:SH_flakes} roughly follows the theoretical sinc-like behaviour, but there is a significant number of outliers. 

We propose three possible reasons (or a combination of those) for the measured differences in SH response: i)~3R polytypism, ii)~stacking faults and iii)~metallic phase changes. For a detailed discussion, see Supporting Information.

In conclusion we have demonstrated TMDC high-index resonators for nonlinear conversion enhancement. 
The resonance behavior of the bulk induced TH suggests the mode to be anapole-like. 
The assembly of our WS\textsubscript{2} resonators into an array reveals the SH emission to originate either from the bulk or from the surface of the resonators, depending on variations in the SH response of the used bulk material. 
A statistical analysis indicates that most bulk flakes offer a small SH signal. 
As those flakes are used for fabricating resonators, we find that SH emission arises from the resonator surface. 
However, material outliers exhibiting an increased SH response can be utilized for resonators that show strong SH emission from the bulk of the resonators. 
We hypothesize that these material outliers are connected to stacking faults and/or polytypism. 
Clearly, further studies on material induced variations of SH generation in bulk TMDCs are needed.

Nevertheless, despite their presumed low bulk nonlinearity, we demonstrated the possibility to utilize bulk TMDCs for building resonant devices which exhibit highly efficient SH generation. 
The potential of our approach becomes clear as one compares the SH intensity of a WS\textsubscript{2} monolayer (integrated SH intensity $=10^4$) to outliers with bulk thicknesses between $100$ and $200\,{\rm nm}$ (integrated SH intensity $=3\times 10^3$), cf. Fig.~\ref{fig:SH_flakes}c. 
By preselecting or engineering those highly efficient bulk flakes and patterning them into resonators, a resonantly enhanced SH intensity [see Eq.~\eqref{eq:enhancement}] of $f_{\rm SH}\times 3 \times 10^3 \approx 15\times 10^4$ could be achieved. 
Therefore, considering a unit area of the nonlinear material, resonantly enhanced SH generation from bulk TMDCs can be more than one order of magnitude more efficient than SH generation from the same TMDC in its monolayer form. 
Additionally the emitted SH power can further be enhanced by maximizing the overlap integral in Eq.~\eqref{eq:overlap}~\cite{Carletti15, Gigli19}. Practically this means that the resonator geometry needs to be optimized for maximal SH power~\cite{Koshelev20}.
Building on these results, preselecting and engineering TMDC bulk materials for maximal SH response, appears as an exciting venue for future research. Patterning those SH enhanced TMDCs into resonator arrays presents a profound opportunity for creating highly efficient nonlinear TMDC meta-surfaces which could be used for LIDAR~\cite{Busschaert19} and holographic imaging applications~\cite{Gao18}.

\section*{Methods}

\subsection*{Electrodynamic Simulations}
Numerical simulations are carried out with the RF module of COMSOL Multiphysics. A single WS\textsubscript{2} disk is defined on top of a glass substrate ($n = 1.52$), whereas the disk is surrounded by air ($n = 1$) in the other half space. 
For the dielectric function of WS\textsubscript{2} we use measured (in-plane) and calculated (out-of-plane) values~\cite{Verre19}, which we linearly extrapolate towards higher wavelengths. 
The simulations are carried out in two steps: 
First, the resulting electric field $\bm{E}_{\rm bkg}(\bm{r})$ of the geometry without disks is simulated under plane wave illumination. 
In a a second step, the disks are included in the simulation and starting from $\bm{E}_{\rm bkg}(\bm{r})$ the targeted field $\bm{E}_{\rm disk}(\bm{r})$ is simulated. 
The two half spaces (above and below the glass-WS\textsubscript{2} interface) are surrounded by perfectly matched layers (PMLs) to exclude back-reflection. 
In addition to the scattering cross-section we calculate the internal electric energy, see Fig.~\ref{fig:anap}, by integrating the total internal electric field over the entire scattering volume,
\begin{equation}
    W_{\rm int} = \frac{\epsilon_0}{2} \int \bm{\epsilon}_{\text{ex}}\left|\pmb{E}\right|^2dV,
\end{equation}
where $\bm{\epsilon}_{\text{ex}}$ is the tensor of the absolute permittivity at the excitation wavelength.

\subsection*{Sample Fabrication}
Glass cover slides with markers are prepared by ultrasonication in acetone with subsequent rinsing in isopropanol and deionized water. 
To remove all organic residues and to roughen the surface for better adhesion, the slides are then plasma-cleaned in 100 W oxygen plasma for 5 minutes. 
Now a sticky tape with exfoliated WS\textsubscript{2} (bulk crystal provided by 2Dsemiconductors.com) is stamped on top of the glass slide and carefully removed. 
In the next step, suitable bulk material samples which sticked to the glass slide are identified via bright and dark field microscopy and their thickness is determined via an AFM measurement. 
The samples are then spin-coated with a conductive polymer (Espacer 300Z, 3000 rpm for 30s with 6s ramp) and baked at 115\textdegree C for 5 minutes. 
For nano-patterning we utilize a Gallium FIB system. 
The single disks are milled in a two step process: 
First, disks with a radius larger than the target radius are milled with a beam current of 7 pA. 
This higher beam current (large FIB radius) is used for a faster milling-rate.
Second, a smaller beam current (small FIB radius) of 2 pA is employed for the final structure. 
The acceleration voltage is maintained at \SI{30}{\kilo\volt} for every milling step. 
Regarding the arrays, the sample preparation before patterning is maintained as for the individual disks. 
To exclude any nonlinear signal from surrounding material, a WS\textsubscript{2} area ($\gtrsim 1/e^2$ intensity diameter of the excitation beam) enclosing the array is removed.
This is achieved by milling (with a 18 pA current) a square of \SI{10}{\micro\meter} x \SI{10}{\micro\meter} with a central exclusion area that is reserved for the array. 
Afterwards four point stars are milled (7 pA) into the left over WS\textsubscript{2} area, such that WS\textsubscript{2} pillars remain with the shape of octagons. 
In the last step disks with the final design radius are milled with a current of 2 pA. 
In contrast to individual disks, which are milled sequentially, arrays are milled in parallel. 
After patterning, the Espacer is washed off by rinsing the sample with deionized water for 2 minutes.

\subsection*{Nonlinear Measurements}
Performing our nonlinear measurements we use the IR output of an optical parametric oscillator (Coherent Mira-OPO) that provides 200 fs pulses at a repetition rate of 76 MHz. 
To exclude any pump contributions we spectrally filter the laser (Semrock BL 1110LP, Chroma HHQ940LP). 

The excitation and emission wavelengths are separated by a dichroic beamsplitter (DMSP950R by Thorlabs). 
Further spectral selection is implemented via optical filters (770SP, 650/60BP for SH generation, 561SP for TH generation at \SI{1310}{nm}, 390/40BP for TH generation at \SI{1200}{nm}, all filters by Semrock) in front of the detectors. 
The emitted optical signal is either sent on an avalanche photo detector (MPD PDM 50), on a spectrometer (Acton SP2300 with Pixis100 CCD) or on an EMCCD camera (Acton Photon Max 512). 
Different lens arrangements are used for real space and BFP imaging. 

For confocal scanning experiments (Fig.~\ref{fig:TH}, \ref{fig:SH} and \ref{fig:SH_flakes}), the beam is directly focused onto the sample using an oil immersion objective (1.3 NA Plan Fluor Nikon). 
Due to underfilling, the resulting beam has a $1/e^2$ intensity radius of (\SI{600}{} $\pm$ \SI{30}{}) \SI{}{\nano\meter}. Thus, it is larger than the maximal disk radius of \SI{400}{\nano\meter}. 
The average excitation power on the sample is measured to be (\SI{570}{} $\pm$ \SI{24}{}) \SI{}{\micro\watt} for the \SI{1310}{nm} excitation wavelength. The average power for the \SI{1200}{nm} excitation wavelength is (\SI{593}{} $\pm$ \SI{25}{}) \SI{}{\micro\watt}. The integrated SH intensity in Fig.~\ref{fig:SH_flakes}c is calculated by determining the area under the recorded SH spectra.

For large area illumination (Fig.~\ref{fig:SH_array}), the beam is sent through a lens (\SI{400}{mm} focal length, AC254-400-C-ML by Thorlabs), that enables us to focus the laser onto the BFP of the objective. The resulting $1/e^2$ intensity diameter in the sample plane is roughly \SI{10}{\um}. 
For these measurements, the average power on the sample is \SI{1.1}{\milli\watt}.

To exclude the possibility that the reported varying SH bulk material responses are due to implanted ions or redistributed material (both from FIB process) or the \mbox{Espacer}, we conduct various tests. 
First, we find that the milling process generally reduces the SH signal (probably due to destruction of the underlying crystal), but does lead to a non-negligible increase in TH signal. 
Thus we cannot exclude that our TH measurements receive significant contributions from implanted ions. 
Second, redistributed material contributes at maximum \SI{50}{\percent} to the total SH signal. 
Third, Espacer, when not washed away by deionized water, leads to a maximum SH signal increase of \SI{56}{\percent}. All these possible influences cannot explain the 7 fold SH signal difference between bulk material 1 and 2.
Thus, by exclusion and supported by the measurements displayed in Fig.~\ref{fig:SH_flakes}c, we are left with the pure bulk nonlinearity as the reason for the observed SH signal difference.

\begin{acknowledgments}
This work was financially supported by the Swiss National Science Foundation (grant no. 200021\_165841). 
The authors thank E. Bonvin and N. Lassaline for valuable input during sample fabrication, and S. Papadopoulos for providing support during data analysis. 
We also would like to thank Dr. M. Timofeeva and F. Richter for providing help with simulations. 
We thank C. Gigli for enlightening discussions on quasinormal resonator modes.
We thank Prof. A. Morpurgo and N. Ubrig for helpful discussions regarding TMDC crystal properties.
Furthermore, the authors acknowledge the use of the facilities of the FIRST center of micro- and nanoscience and of ScopeM at ETH Zürich. 
\end{acknowledgments}

\section*{Supporting Information}
The Supporting Information includes:\\
sample fabrication; spectra of TH measurement; wavelength dependence of arrays; single layer dependence of SH signal; reasons for the measured differences in SH response.

\nocite{*}
\providecommand{\noopsort}[1]{}\providecommand{\singleletter}[1]{#1}%



\section*{Supplementary material}

# Supporting Information: TMDC Resonators for Second Harmonic Signal Enhancement

Sebastian Busschaert,[1] René Reimann,[1] Moritz Cavigelli,[1] Ronja Khelifa,[1] Achint Jain,[1] and Lukas Novotny[1]

[1]*Photonics Laboratory, ETH Zürich, CH-8093 Zürich, Switzerland*

## I. SAMPLE FABRICATION

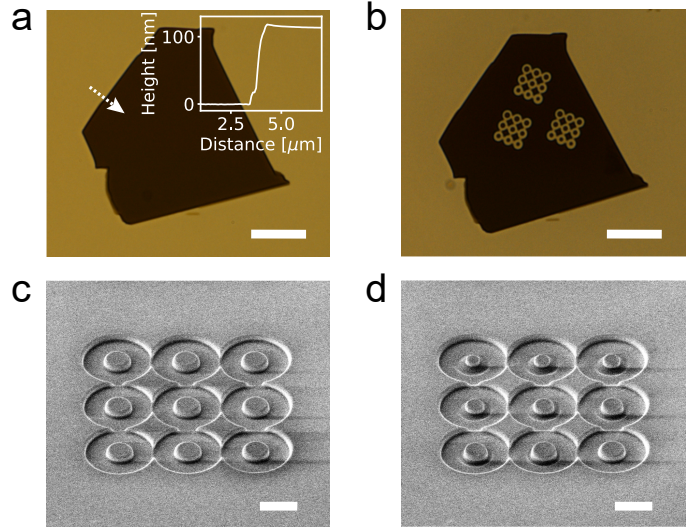

FIG. S 1. Fabrication of single WS$_2$ disks on glass. (a) Transmission microscope image of exfoliated WS$_2$ on glass. The scale bar is 5 µm. The inset shows a line profile of an AFM scan, indicated by the dashed arrow. (b) Transmission microscope image of the same flake after nano-patterning. (c,d) SEM images of the (c) rough and (d) fine FIB processing steps. The scale bar is 1 µm.

In Fig. S1 we provide more details on our sample fabrication, which we also discuss in the Methods section of the main text. Fig. S1a shows an transmission microscope image of exfoliated WS$_2$ bulk material on glass. The high optical contrast between flake and substrate is indicative of the large thickness of the material, which we confirm via an AFM scan (thickness here ~120 nm). Fig. S1b shows the same flake after patterning, now with visible exemptions surrounding the patterned disks. In Figure S1c and d we show the two FIB processing steps that result in the single disk arrays. All milling steps are performed with an ion acceleration voltage of 30 kV. Fig. S1c shows the disks after the first FIB step, where we pattern slightly larger disks with a higher beam current (7 pA). Utilizing larger beam currents generally results in shorter milling times, but goes in hand with a loss of resolution. This is due to the fact that the size of the milling beam spot is proportional to the beam current. In Fig. S1d we show the disks after the second FIB step, for which we use a smaller beam current (2 pA) with higher resolution.

Arrays of disks, see e.g. Fig. S2a,b, are fabricated with two important distinctions. First, a fabrication step is added (17 pA) before the rough milling of the disks (7 pA). This 17 pA milling step removes the TMDC bulk material surrounding the array in order to avoid SH background as we probe with a large area beam spot ($1/e^2$ intensity diameter of about 10 µm). Second, each milling step is performed in parallel. Instead of writing disk after disk (which is done for the single disk samples, see e.g. Fig. S1d), all disks are milled simultaneously. Without parallel writing, edges would form in a sequential manner and influence subsequently milled disks, thus leading to an asymmetric array.

For all samples, we chose milling times such, that we are sure to have removed all WS$_2$ material between the disks and the remaining bulk flake. This results in over-milling (i.e. milling into the glass substrate) of ~100 nm for the individual disks and ~200 nm for the arrays. We conduct electrodynamic simulations of over-milled structures which show no significant change in the spectral position of the disk resonances compared to the simulations discussed in the main text. The same conclusion is reached in [1], which studies the influence of a 300 nm glass crown (on top of a WS$_2$ resonator) on the there described resonator's scattering efficiency.

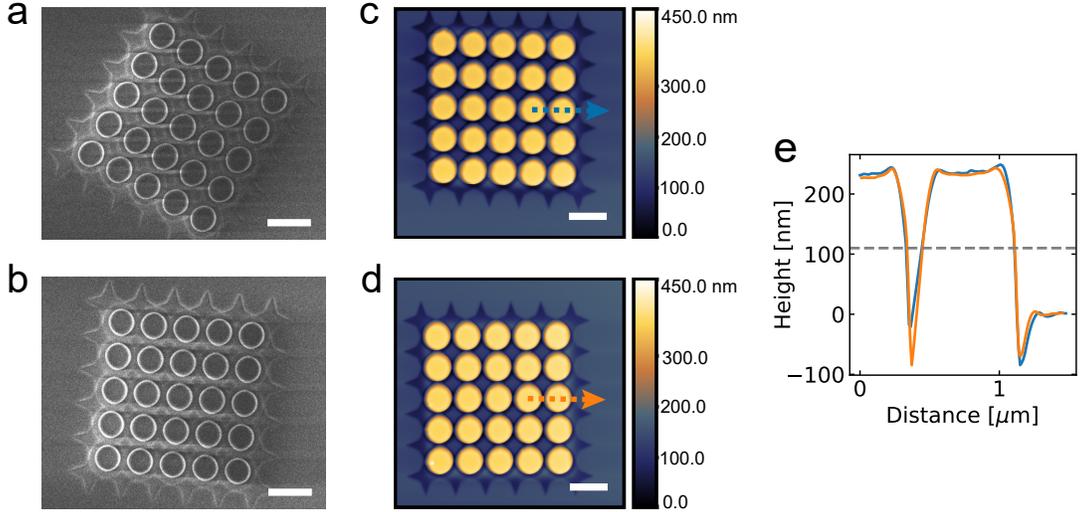

FIG. S 2. Fabrication reproducibility of WS$_2$ disk arrays. (a,b) SEM images of two patterned arrays with identical design radius (300 nm) and pitch (780 nm) on different flakes. The scale bars are 1 µm. (c,d) AFM images of two patterned arrays with the identical design radius (300 nm) and pitch (780 nm) on different flakes with the same thickness (123 nm). The scale bars are 1 µm. (e) Cross sections of the AFM scans in (c) and (d), along the respective colored dashed lines. The grey dashed line marks the bottom interface of the WS$_2$.

In order to to demonstrate the reproducibility of our fabrication process, we present a post-characterization of fabricated resonator arrays in Fig. S2. To this end we image two different arrays, which have identical design parameters (Fig. S2a,b). The disk shape is circular (asymmetry of disk less than 5 %) and comparing the respective images reveals a close match between the two arrays. Within the image resolution we estimate the difference in radius ($|r_{array,1} - r_{array,2}|/r_{array,2}$) to be less than 5 %. Apart from the radius, the height of the resonators is an important geometrical quantity. To study the height profile we conduct AFM scans on arrays of resonators. Here we investigate equally processed arrays on two flakes with identical height of 123 nm, see Fig. S2c and d. The slight asymmetry of the disks arises from the asymmetric AFM tip. Taking cross sections in Fig. S2e and comparing the total height differences between the top of the pillars and the substrates we find that the milling depth is reproducible. Additionally, the disc radius at the bottom interface of the flake (dashed grey line) is reproducible, too.

## II. SPECTRA OF TH MEASUREMENTS

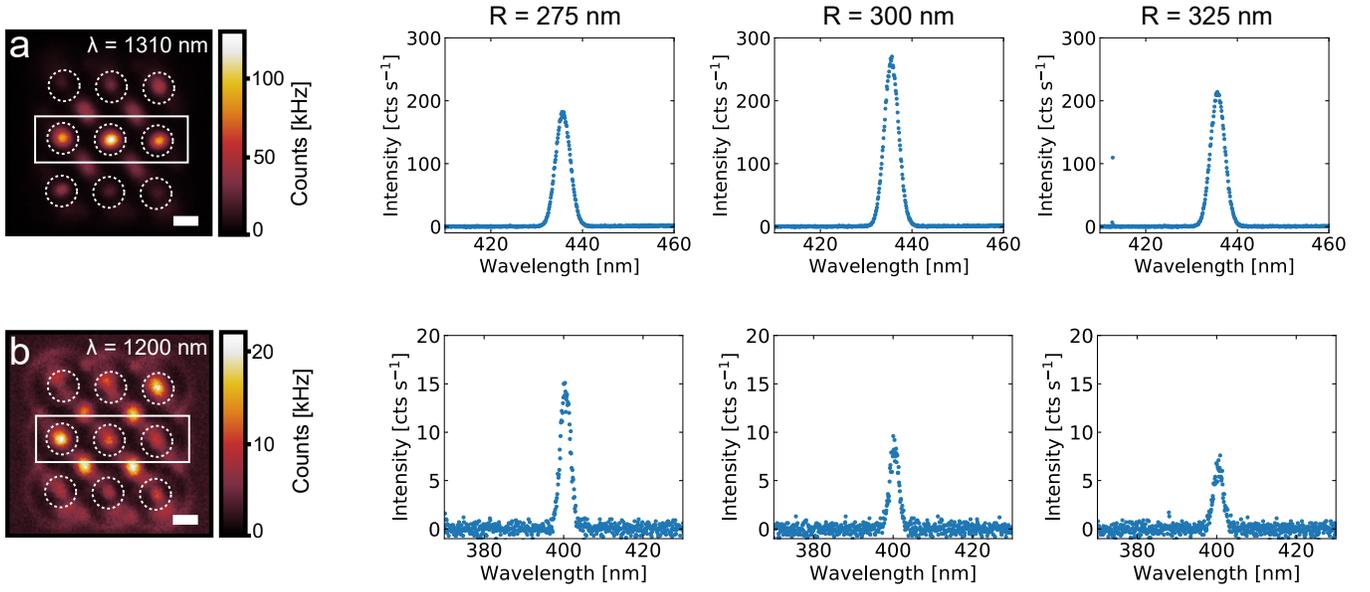

FIG. S 3. TH spectra corresponding to the intensity maps in Fig. 2c and d in the main text. (a,b) TH intensity maps (identical to Fig. 2c and d) for excitation wavelength of (a) 1310 nm and (b) 1200 nm, with TH spectra collected from the disks in the respective white boxes.

In Figure S3 we display spectra of the TH emission which belong to the TH intensity maps that are displayed in the main text in Fig. 2. The TH intensity maps are measured with an avalanche photo diode (APD) which integrates the signal. Thus, here we show that the resonance behavior is given by the coherent TH emission and not due to possibly present incoherent background. Indeed we find that the respective resonance behaviour of the disks within the white boxes is confirmed in the recorded spectra.

## III. WAVELENGTH DEPENDENCE OF ARRAYS

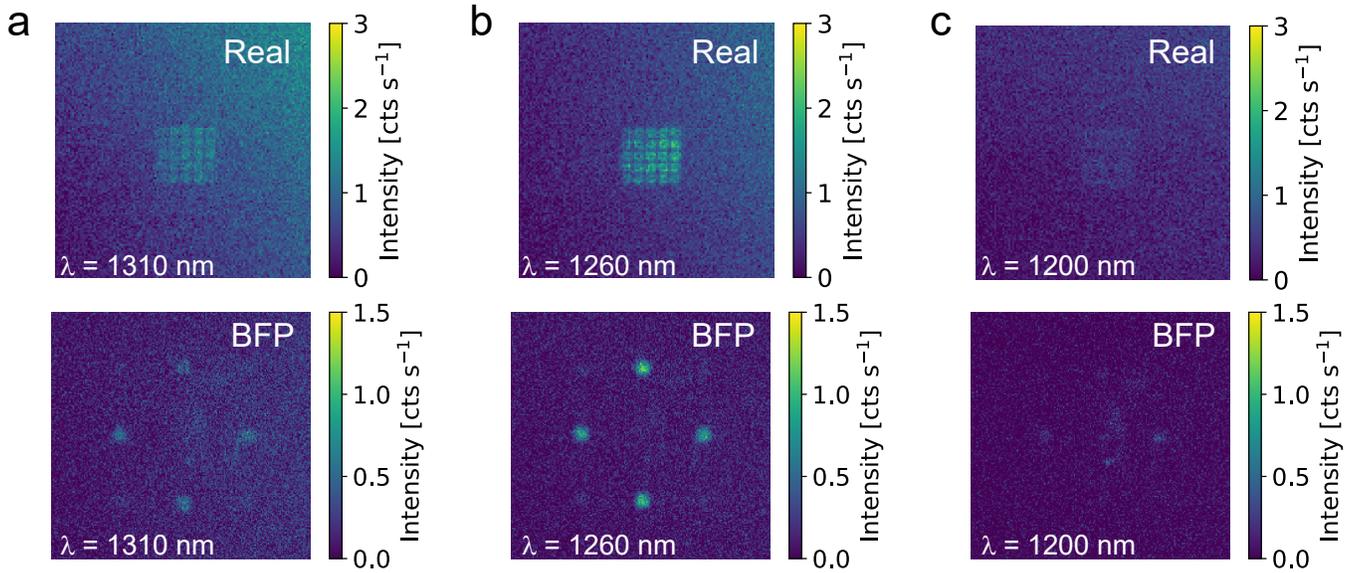

FIG. S 4. Wavelength dependence of SH signal for device 2. Real and BFP images for an excitation wavelength of (a) 1310 nm, (b) 1260 nm and (c) 1200 nm. For comparison, the false color plots of the intensity are all scaled to the same value intervals (real and BFP respectively). The average excitation powers are kept constant with less than 5 % variation.

Here we study the SH response of device 1 and 2 for different excitation wavelengths. We start with device 2, as this device is more representative of other measured arrays. To see whether we can observe a zeroth order emission lobe similar to device 1, we vary the excitation wavelength. The results are shown in Figure S4. As we tune the wavelength to 1260 nm (Fig. S4b), we observe a signal increase compared to excitation at 1310 nm, but still no zeroth order emission lobe. As we tune the wavelength even further to 1200 nm (Fig. S4c) the SH signal decreases again. The recorded data is not normalized to the transmission function of the measurement setup, but we note that our optical components show less losses as the wavelength is tuned towards the visible. The incident power and the integration times were kept constant for all wavelength changes. The intensity changes from a to c in Fig. S4, therefore indicate that we have mapped out the array's resonance. However, the absence of a zeroth order emission lobe suggests that the observed resonance is not primarily anapole-like. As the spectral peak of the SH resonance is blue-detuned with respect to the observed anapole resonance at the fundamental wavelength, we speculate that the resonance of device 2 has an electric dipole character [2–4].

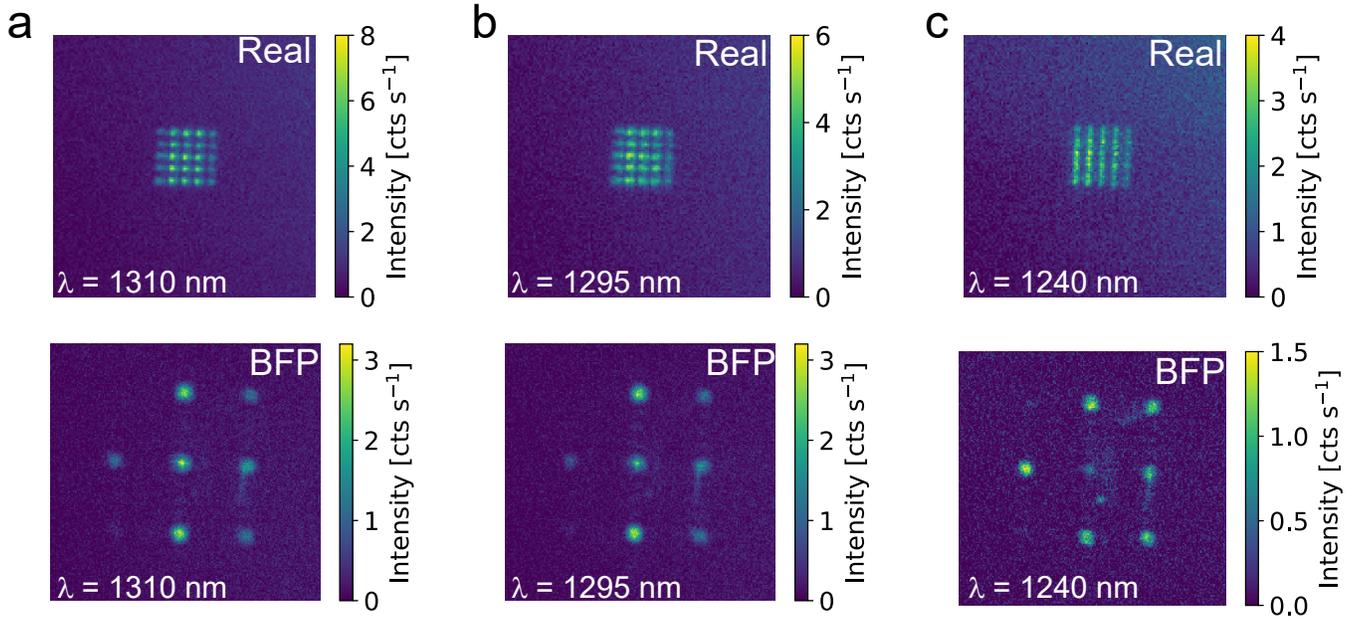

FIG. S 5. Wavelength dependence of SH signal for device 1. Real and BFP images for an excitation wavelength of (a) 1310 nm, (b) 1295 nm and (c) 1240 nm. The excitation powers for 1310 nm and 1240 nm are kept constant within less than 5 % variation. The excitation power for 1295 nm is 9 % higher.

Next, we analyse the behavior of device 1. As we tune the excitation wavelength towards the blue (see Figure S5b), we observe an overall decrease in counts, both in real space and in the BFP. In addition the real space images suggest that sharp features which are visible at 1310 nm get blurred. In the BFP the zeroth order SH emission lobe (bulk emission) fades with decreasing excitation wavelength. This finding agrees with the interpretation that, by tuning the wavelength, we are not resonant any more with the anapole-like resonator mode which enhances SH bulk emission. Consequently, the BFP image in Fig. S5c (at a far off resonant excitation wavelength of 1240 nm) is similar to the BFP images of device 2, shown in Figure S4.

## IV. SINGLE LAYER DEPENDENCE OF SH SIGNAL

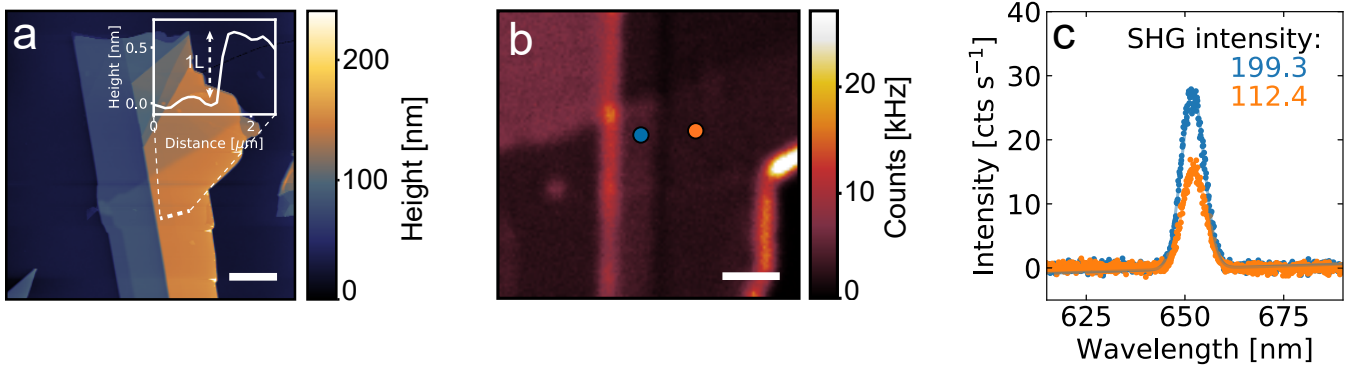

FIG. S 6. Single layer dependence of SH signal in bulk TMDCs. (a) AFM scan of a pristine $WS_2$ bulk material with a thickness (about 125 nm) comparable to the bulk material of devices 1 and 2. The scale bar is 5 μm. The inset shows an AFM scan along a direction which is indicated by the white dashed line. (b) Corresponding SH map with excitation wavelength at 1310 nm. Blue: $n$ layers. Orange: $n+1$ layers. The scale bar is 2 μm. (c) Spectra taken at the points indicated in (b). The numbers are the areas under the Gaussian fits.

The bulk materials of device 1 and 2 are both determined to be about 123 nm thick. Here we exclude an odd/even layer number dependence as an origin for the observed factor 7 difference in SH response between device 1 and 2. For thin layers, as described in [5, 6] an odd layer number leads to a non-centrosymmetric crystal (SH allowed) and an even layer number to a centrosymmetric crystal (SH forbidden). Previous studies in literature have not looked at the same distinction for thick flakes which were considered to be bulk with layer number independent optical properties. In Figure S6 we display a representative bulk measurement with a single layer step in the material. Measurements of other flakes with single layer steps showed comparable behavior. Fig. S6a shows an AFM scan of a flake with a thickness of about 125 nm. An AFM line scan (white dashed line, inset) indicates a single layer step. In the SH map in Fig. S6b we indeed see a distinct step in SH signal that can be assigned to the height difference of a single layer. To exclude any background contribution in the maps we take spectra at the indicated points and plot them in Fig. S6c. The areas under the fitted curves reveal a signal difference of 70 %.

## V. REASONS FOR THE MEASURED DIFFERENCES IN SH RESPONSE

In the main text we find that the measured SH response strongly depends on the used WS$_2$ material flake. Here we discuss three possible reasons (or a combination of those) for this difference in SH response: i) 3R polytypism, ii) stacking faults and iii) metallic phase changes.

i) Polytypism is the existence of multiple phases within one crystal [7]. The most widely studied phase in TMDCs is the 2H phase, which exhibits alternating high and low SH responses for odd and even layer number, respectively [8, 9]. However, there exists a different phase, called 3R, that shows an increase in SH response with increasing layer number [9]. This increase is due to the different alignment of individual layers with respect to each other, which can result in net constructive interference. The here used crystal of WS$_2$ is characterized as 2H by its vendor. On the other hand, studies of polytypism in MoS$_2$, namely the existence of the 2H and 3R phase in the same crystal [10], link polytypism with an increase of SH generation in bulk samples [11]. The TMDC in these studies is not WS$_2$, but the similarities among TMDCs make it likely for WS$_2$ to also exhibit this behavior.

ii) Stacking faults are considered to be a crystallographic defects with disordered crystal planes (planar defect) [12]. In 2H TMDCs the stacking sequence is ABAB, with interlocking chalcogenide atoms. However, a displacement/rotation of a layer can lead to a discontinuity (ABCAB) and thus generate an enhanced SH response due to symmetry breaking or constructive interference between differently rotated layers. The possibility of stacking faults was confirmed to us by the manufacturer of the provided WS$_2$.

iii) Another explanation for the observed strong variations in the SH bulk response is the appearance of a metallic phase, which has been shown to exhibit high SH signals [13]. However, these phases have to be engineered in specific ways, so an appearance in our purchased crystals seems unlikely.

---



\end{document}